\documentclass[prb,aps,twocolumn,superscriptaddress,showpacs,amsmath,amssymb,floatfix]{revtex4}
\usepackage{graphicx}
\newcommand{\be}{\begin{equation}}
\newcommand{\ee}{\end{equation}}
\newcommand{\xx}{{\mathbf x}}
\newcommand{\kk}{{\mathbf k}}
\newcommand{\qq}{{\mathbf q}}
\newcommand{\eq}[1]{(\ref{#1})}

\newcommand{\im}{{\rm Im}}

\begin{document}
\title{The parametric oscillation threshold of semiconductor
  microcavities in the strong coupling regime} 
\author{Michiel Wouters}
\affiliation{BEC-INFM and Dipartimento di Fisica, Universit\`a di Trento, 
1-38050 Povo, Italy}
\affiliation{TFVS, Universiteit Antwerpen, Universiteitsplein 1,
2610 Antwerpen, Belgium}
\author{Iacopo Carusotto}
\affiliation{BEC-INFM and Dipartimento di Fisica, Universit\`a di Trento, 
1-38050 Povo, Italy}

\begin{abstract}
The threshold of triply resonant optical parametric oscillation in a
semiconductor microcavity in the strong coupling regime is
investigated.
Because of the third-order nature of the excitonic nonlinearity, a variety of
different behaviours is observed thanks to the interplay of parametric
oscillation and optical bistability effects.
The behaviour of the signal amplitude and of the quantum fluctuations in
 approaching the threshold has been characterized as a function of the
 pump, signal and idler frequencies.
\end{abstract}

\maketitle

\section{Introduction}

Triply resonant optical parametric oscillation
(OPO)~\cite{OPO_general,OPO_special} has been recently 
observed~\cite{stevenson,houdre,baumberg}  
in semiconductor microcavities in the strong coupling
regime~\cite{microcavity_review1,microcavity_review2},  
and has attracted a good
deal of attention from the point of view of both fundamental physics
and possible technological
applications.
The peculiar dispersion relation of polaritons in the strong coupling
regime allows to simultaneously satisfy the resonance condition for the pump, 
signal and the idler modes.
Together with the enormous value of the excitonic nonlinearities, the possibility
of easy phase matching results in a low threshold intensity, making these systems 
very promising candidates for low-power OPO applications.

A complete theoretical description of the OPO dynamics of such systems
is not only very important in view of the optimisation of the device
operation, but also deserves a certain interest from the point of
view of nonlinear dynamics as many interesting phenomena can occur due
to the interplay of optical bistability and parametric
oscillation~\cite{iac-superfl,whittaker05}, and to the nontrivial spatial field
dynamics in the transverse plane~\cite{whittaker_last,pattern}. 

As shown by several theoretical papers that have appeared on the
subject, a rather
complex phenomenology is found already at the level of the three-mode
approximation, where the classical
nonlinear optical wave equation is projected onto the three pump, 
signal and idler modes~\cite{whittaker05,whittaker01,ciuti-rev,gippius}.
Available experimental data appear to confirm this point: in 
particular, both continuous~\cite{baumberg,dasbach} and
discontinuous~\cite{baas} behaviours have been experimentally shown
for the signal intensity in the neighborhood of the threshold point.
Although some analogies have been drawn with what is known about
$\chi^{(2)}$ OPO dynamics in standard passive
media~\cite{OPO_general,OPO_special,TROPO,chi2OPO,lugiato}, no
complete investigation has appeared yet for the case of semiconductor
microcavities in the strong coupling regime, neither from the
experimental nor from the theoretical points of view.

The optical nonlinearity of the microcavity system under investigation
originates from collisional exciton-exciton interactions
and is therefore of the $\chi^{(3)}$ type.
This means that it not only provides the parametric interaction
necessary for the parametric oscillation, but is also responsible for
significant mean-field frequency shifts of the modes. 
This makes the nonlinear dynamics of the mode amplitudes much richer
than in $\chi^{(2)}$ OPOs~\cite{lugiato}. Pioneering theoretical
work in this direction has recently appeared in
Ref.~\onlinecite{whittaker05}. 

The purpose of the present paper is to provide a systematic and
quantitative study of the OPO threshold in semiconductor microcavities
in the strong coupling regime.
Depending on the pump laser frequency, and the signal, pump and idler
mode frequencies, several regimes are to be distinguished,
where the system behaviour is radically different.

The paper is organized as follows: our model of the microcavity is
introduced in Sec.\ref{sec:model}.
Optical limiting and optical bistability in the pump only solution are
discussed in Sec.\ref{sec:pump_only}. 
General concepts about the stability of the solution with respect to
pump-only and to parametric instabilities are given in
Sec.\ref{sec:OPO}.
The following Secs.\ref{sec:angle}-\ref{sec:magic_angle} are devoted to
characterize the parametric threshold as a function of
the incident pump angle, the internal and the incident
intensities and to find the optimal choice to minimize the threshold
intensity.
Quantitative estimations are provided in Sec.\ref{sec:quant}, where
a comparison is made with other realizations of OPOs based on passive
$\chi^{(2)}$ and $\chi^{(3)}$ materials. 
The kind of bifurcation at the onset of the OPO emission is investigated 
in Sec.\ref{sec:bif}.
Depending on whether the pump-only solution is in the optical limiter or
in the optical bistability regimes, parametric emission is shown to set
in either in a continuous or in a discontinuous way.
The close relationship between the nature of the instability point and
the behaviour of the quantum fluctuations as the threshold is approached
is pointed out in Sec.\ref{sec:fluct}.
Conclusions are finally drawn in Sec.\ref{sec:conclusions}.

\section{Polariton model \label{sec:model}}

\begin{figure}[htbp]
\begin{center}
\includegraphics[width= \columnwidth,angle=0,clip]{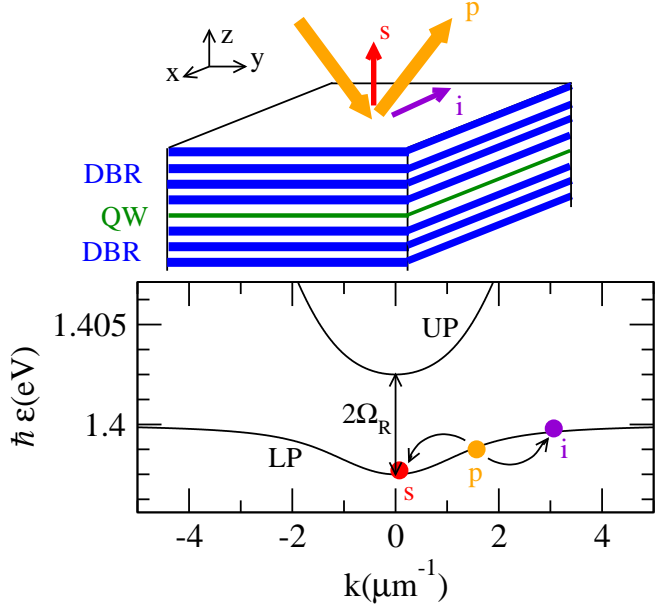}
\caption{
Upper panel: sketch of the microcavity system and of the parametric
process under consideration.
Lower panel: lower polariton (LP) and upper polariton (UP)
dispersion at linear regime. The cavity photon dispersion is
$\omega_{C}(k)=\omega_{C}^0\,\sqrt{1+{k^2}/{k_z^2}}$
  with
$\hbar\omega^0_C=1.4$~eV and $k_z=20\,\mu\rm m^{-1}$.
The exciton dispersion is flat and resonant with the $k=0$ cavity mode
$\omega_X=\omega_C^0$. The exciton-photon Rabi coupling is
$\hbar\Omega_R=2.5$~meV. The dots indicate the signal, pump and
  idler modes, the arrows show the triply resonant parametric process under
  investigation.}
\label{fig:sketch}
\end{center}
\end{figure}

A sketch of the physical system under investigation is shown in 
Fig.\ref{fig:sketch}: a planar DBR (Distributed Bragg
Reflector) semiconductor microcavity containing a few quantum wells
strongly coupled to the cavity mode.
The elementary excitations of this system consist
of exciton-polaritons, i.e. coherent superpositions of cavity photons
and excitons.
Polaritons combine the very strong $\chi^{(3)}$ optical nonlinearity
originating from exciton-exciton collisional interactions to the
peculiar dispersion relation as a function of the in-plane wavevector
$\kk$ that is shown in Fig.\ref{fig:sketch}: these facts make them extremely
well suited for triply-resonant optical parametric oscillator
applications, as it has indeed 
been experimentally demonstrated in recent
years~\cite{stevenson,houdre,baumberg}.

A mean-field description of the cavity-polariton field dynamics can
be developed in terms of a nonlinear wave equation with
a third-order nonlinearity~\cite{ciuti-rev,iac-superfl}.
Under the assumption that the dynamics takes place in the lower
polariton branch and the population of the upper polariton branch
remains negligible, the theoretical description can be simplified by
restricting it to the lower polariton only.
For the sake of simplicity, we shall focus our attention on the case of
a circularly polarized pump beam.
As the circular polarization of the polariton field is preserved 
by the nonlinear interactions and the longitudinal-transverse
splitting~\cite{kavokinpss,langbein} is much smaller than both the linewidth
$\gamma$ and the nonlinear interaction energy, the circular polarization
is almost completely transferred to the signal and idler beams.

Under these assumptions, the polariton dynamics can be
written in terms of a nonlinear wave equation for a single-component
$k$-space polariton field $\psi_{LP}(\kk)$:
\begin{multline}
i \frac{d}{dt}\psi_{LP}(\kk) =\left[\epsilon(k)
-i\frac{\gamma(\kk)}{2}\right] \psi_{LP}(\kk)
+F_{p}(\kk)\,\, e^{-i\omega_{p}t}  \\
+  \sum_{\qq_1,\qq_2} g_{\kk,\qq_1,\qq_2}\, \psi^*_{LP}(\qq_1+\qq_2-\kk)
\, \psi_{LP}(\qq_1)\, \psi_{LP}(\qq_2).
\label{eq_mot}
\end{multline}
The field $\psi_{LP}(\kk)$ is here normalized in such a way that
its square modulus  
$|\psi_{LP}(\kk)|^2$ equals the number of polaritons with momentum $\kk$ 
per unit area. 
$\epsilon(\kk)$ is the dispersion relation of the lower
polariton and $\gamma(\kk)$ is the momentum-dependent loss rate.
Throughout the present paper, the exciting laser field is taken as a 
monochromatic and continuous wave coherent field at $\omega_p$ with a
plane-wave  spatial profile at $\kk_p$ and a circular polarization.  
The driving amplitude $F_p(\kk)$ can be related to the incident power
density $I_{inc}$ by 
using the input-output formalism~\cite{gardiner,input-output,baas}:
\begin{equation}
F_p(\kk)=\delta_{\kk,\kk_p} C(\kk_p) 
\sqrt{\frac{\gamma_{rad} I_{inc}}{N_{tr}\hbar\omega_p}}.
\label{def:F_p}
\end{equation}
$\gamma_{rad}$ is here the radiative decay rate of the cavity-photon 
due to the finite mirror transmittivity;
the parameter $N_{tr}$ specifies whether the cavity is a single-sided
cavity with a 
perfectly reflecting back mirror ($N_{tr}=1$), or a symmetric cavity
with equal transmission through both the front and back mirrors
($N_{tr}=2$).

The third-order nonlinear interaction term takes into account
exciton-exciton collisional interactions. As the wavevectors involved
in the present discussion are much smaller than the inverse excitonic
radius, the exciton-exciton coupling constant in a single quantum well
can be approximated by a momentum-independent ${\bar g}$.
If $N_{QW}$ quantum wells are present in the cavity, identically
coupled to the cavity mode, the bright excitonic excitation is
delocalized over all of them and the effective excitonic coupling
constant is $g={\bar g}/N_{QW}$.
In the polaritonic basis, a non-trivial momentum dependence appears 
via the Hopfield coefficients $X(\kk)$ and $C(\kk)$ quantifying the
excitonic and cavity-photonic components of the lower polariton:
\begin{equation}
g_{\kk,\qq_1,\qq_2}=g\, X^*(\kk)\, X^*(\qq_1+\qq_2-\kk)\,
X(\qq_1)\, X(\qq_2). 
\end{equation}
Although no conclusive experimental nor theoretical analysis has been
reported yet, the theoretical prediction $\hbar {\bar g}\approx
1.5\times10^{-5}\,\textrm{eV}\, \mu \textrm{m}^2 $ based on the Born
approximation~\cite{Born_approx} appears to be in reasonable agreement
with available experimental 
data~\cite{microcavity_review1,microcavity_review2}. 

In order to focus our discussion of the basic OPO dynamics, we shall not
consider here the effect of the disorder: in recent high quality
III-V samples the effect of the disorder can in fact be weak enough for
it to be 
neglected on the scale of the polaritonic linewidth. In this case, it is
legitimate to approximate the mode eigenfunctions as plane waves.
On the other hand, the disorder is much stronger in II-VI
samples, where it has been shown to have dramatic consequences on
polariton BEC~\cite{Pol-BEC}.
These effects are highly non-trivial already at
equilibrium~\cite{fisher} and are expected to become even more complex
because of the interplay with the nonlinear dynamics: the complete
analysis of them goes far beyond the scope of the present paper and is
left to future work.

To conclude the section, it is interesting to note that the
applicability range of the wave equation \eq{eq_mot} is not 
limited to semiconductor planar OPOs, but can be extended to 
describe other setups, e.g. planar cavities containing a slab
of passive $\chi^{(3)}$ material.
In this case, no excitonic resonance exists, and the
polariton reduces to a bare cavity-photon.
As both the coupling to external radiation and the optical
nonlinearity act on the same photonic degree of freedom, one
has simply to set $X=C=1$ and calculate the nonlinear coupling
constant using the nonlinear susceptibility of the medium under
consideration:
\begin{equation}
\hbar g={\mathcal C}\, \chi^{(3)}\,\frac{(\hbar\omega_p)^2}{\epsilon_{\rm lin}^2 d}.
\label{rel:g_chi3}
\end{equation}
The numerical factor ${\mathcal C}$ of order one takes into account
the details of the geometry under investigation.
Typical values of $\chi^{(3)}$ of materials specifically
designed for nonlinear optical applications range up to something of
the order of $10^{-9}\,\textrm{esu}$~\cite{NLO_mat}.
For a $\lambda/2$ cavity, these values correspond to a nonlinear
coupling constant of the order of $\hbar g\approx 5\times
10^{-9}\,\textrm{eV}\,\mu\textrm{m}^2$, orders of magnitude lower
than the value $\hbar {\bar g}\approx 1.5\times10^{-5}\,\textrm{eV}\,
\mu \textrm{m}^2 $ previously mentioned for semiconductor
microcavities in the strong coupling regime.
This explains the present interest of semiconductor microcavities for
low-power nonlinear optical applications, as well as for fundamental
studies of the interplay of nonlinear dynamics and quantum
fluctuations~\cite{coh_length,verger}.

\section{The parametric threshold}

\subsection{The pump only solution\label{sec:pump_only}}

\begin{figure}[tb]
\begin{center}
\includegraphics[width=\columnwidth,angle=0,clip]{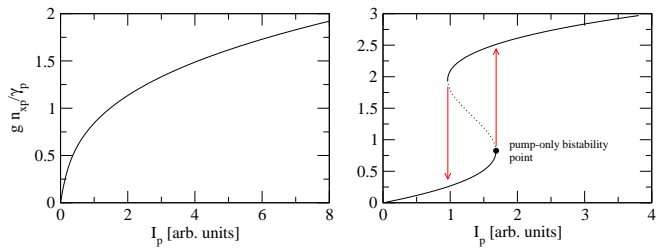}
\caption{
Polariton density in the pump mode as a function of the driving intensity $I_p=|F_p|^2$
for a pump wave vector of $k_p=1.19 \mu\textrm{m}^{-1}$, which
corresponds to $\epsilon_p=1.39845 \,\textrm{eV}$. The left graph is in the optical limiter regime
($\omega_p=\epsilon_p$),  whereas the right one is in the bistability regime
($\omega_p= \epsilon_p+1.5 \gamma_p$). 
The red arrows show the jump in polariton density for an upward (right arrow) and
downward (left arrow) ramp of the laser intensity.
The dotted part of the curve is dynamically unstable.
Same cavity parameters as in Fig.\ref{fig:sketch}.
} 
\label{fig:lim-bist}
\end{center}
\end{figure} 

Among the different $\kk$ modes, only the one at $\kk=\kk_p$ contains 
a source term in its equation of motion \eq{eq_mot}.
An exact solution of the full set of equations of motion \eq{eq_mot}
can therefore be found in the form
\begin{equation}
\psi_{LP}(x,t)=P\, e^{i(\kk_p \xx-\omega_p t)},
\end{equation}
with the amplitude $P$ fixed by the condition:
\begin{equation}
[\epsilon_p-\omega_p-\frac{i}{2}\gamma_p
+g |X(k_p)|^4 |P|^2]P+ F_p=0.
\label{pump_only}
\end{equation}
Here $\epsilon_p=\epsilon(\kk_p)$ is the frequency of the pump mode at
linear regime and $\gamma_p=\gamma(\kk_p)$ the corresponding
linewidth; the effect of the third-order nonlinearity is to 
renormalize the pump mode frequency by a mean-field shift proportional
to the mode excitonic population $n_{xp}=|X(\kk_p)|^2|P|^2$. 
This effect, absent in $\chi^{(2)}$ cavities, is responsible for the
the qualitatively different behaviours~\cite{bistability} that can be
observed depending on the sign of the detuning  
of the pump frequency  $\omega_p$ with respect to the polariton energy
$\epsilon(k_p)$. 

Fig.\ref{fig:lim-bist} shows the excitonic density in the pump
mode $n_{xp}$ as a function of the driving intensity $I_p=|F_p|^2$
which is proportional (but not identical) to the incident laser intensity $I_{inc}$
[See Eq.\eq{def:F_p}].
When the pump frequency is below or close to resonance
$\omega_p-\epsilon(\kk_p)< \sqrt{3}\,\gamma(\kk_p)/2$, we are in the
so-called optical limiter regime, in which the population $n_{xp}$ of
the $\kk_p$ mode monotonically increases as a function of the driving
intensity $I_p$ (left panel). 

For blue-detuned pump frequencies
$\omega_p-\epsilon(\kk_p)>\sqrt{3}\,\gamma(\kk_p)/2$, a positive
feedback of the nonlinearity occurs and hysteretic behaviour can be
instead observed, as shown in the right panel of Fig.\ref{fig:lim-bist}
and experimentally demonstrated in Ref.~\onlinecite{baas-bist}. 
For increasing laser intensity, the pump mode population follows 
the lower branch until its endpoint is reached, and then it jumps to the
upper branch as indicated by the arrow.  
If the driving intensity is later ramped down, the system keeps following
the upper branch until its endpoint, and only here the pump mode
population jumps back to the lower branch.
Hysteretic behaviour is apparent, as the upward and downward jump points
do not coincide.

\subsection{Dynamical stability of the pump-only state\label{sec:OPO}}

As usual in nonlinear dynamical systems, finding a solution is not
sufficient, as one has to verify its dynamical stability.
Optical parametric oscillation, as well as the instability of the
central branch of the hysteresis loop are in fact due to the solution
\eq{pump_only} becoming dynamically unstable.

As the planar cavity supports a continuum of independent modes with
different in-plane wavevectors, dynamical stability of the pump only
solution \eq{pump_only}
has to be checked with respect to perturbations with any wave
vector $\kk_s$:
\begin{multline}
\psi_{LP}(\xx,t)=P e^{-i\omega_p t + i \kk_p  \xx} + u(\kk_s) 
e^{-i[\omega_p+\omega(\kk_s)] t + i \kk_s x} \\
+ v^*(\kk_s) e^{-i[\omega_p-\omega^*(\kk_s)] t + i (2 \kk_p-\kk_s) \xx}.
\label{psi_lin_po}
\end{multline}
Substituting this expression in Eq.\eq{eq_mot} and keeping only linear
terms in the fluctuations $u$ and $v$, one gets to the following
eigenvalue problem~\cite{ciuti-lum}
\begin{equation}
L(\kk_s)\,w(\kk_s) = \omega(\kk_s)\,w(\kk_s)
\label{bog_pump_only}
\end{equation}
where the two-component vector $w(\kk_s)=[u(\kk_s), v(\kk_s)]^T$ and the 2x2 
matrix $L(\kk_s)$ is
\begin{widetext}
\begin{equation}
L(\kk_s)=
\left( \begin{array}{cc}
\epsilon_s-\omega_p+2g|X_s|^2 |X_p|^2 |P|^2-i\frac{\gamma_s}{2}&g X^*_sX_iX_p^2 P^2 \\
-gX_sX^*_iX_p^{*2} P^{*2} & -\epsilon_i+\omega_p-2g|X_i|^2 |X_p|^2|P|^2
-i\frac{\gamma_i}{2}
\end{array}\right).
\label{bog_matr}
\end{equation}
The matrix $L(\kk_s)$ couples the fluctuations in the $\kk_s$ and
$\kk_i=2\kk_p-\kk_s$ modes, called in the following the {\em signal} and
the {\em idler} modes.
Short-hand notations have been here introduced to simplify the expressions:
$X_{s,i}=X(\kk_{s,i})$ are the excitonic Hopfield coefficients of the
 signal/idler modes, $\epsilon_{s,i}=\epsilon(\kk_{s,i})$ are the signal
 and idler mode frequencies and $\gamma_{s,i}=\gamma(\kk_{s,i})$ are the
 corresponding loss rates. 
Dynamical stability is ensured if the imaginary parts of all
eigenvalues of $L(\kk_s)$ are negative $\im[\omega_{\pm}(\kk_s)]<0$ for all
wavevectors $\kk_s$.
These can be written as:
\begin{equation}
\im\{\omega_{\pm}(\kk_s)\} = -\frac{\gamma_s+\gamma_i}{4}\pm
\im\left\{\sqrt{[(\epsilon_{si}-\omega_p)+g(|X_s|^2+|X_i|^2) n_{xp}
-i\frac{\gamma_s-\gamma_i}{4}]^2- g^2\,|X_s|^2 |X_i|^2 n_{xp}^2 }\right\},
\label{imlambda}
\end{equation}
\end{widetext}
Note that the pump mode frequency $\epsilon_p$ does not directly
appear in the expression \eq{imlambda} of the eigenvalues, but it is
only indirectly involved via the pump-only solution \eq{pump_only}, which 
fixes $n_{xp}$.
The frequencies of the signal and idler modes are involved in
\eq{imlambda} only via their average value
$\epsilon_{si}=[\epsilon_s+\epsilon_i]/2$. 

Two kinds of physically distinct instabilities can arise.
A {\em single-mode} instability arises when the equation of motion for
the pump mode alone -- neglecting all interactions with other modes --
is dynamically unstable.
This instability is found when $L(\kk_s)$ has an eigenvalue with a
positive imaginary part for $\kk_s=\kk_p$.
As in this case $\kk_i=\kk_s=\kk_p$, this instability involves the
$\kk_p$ mode only, and for this reason it is called {\em single-mode}.
It is easy to verify~\cite{iac-superfl} that the pump-only solution \eq{pump_only} is
single-mode unstable in the central branch of the bistability curve
(marked with a dotted line in Fig.\ref{fig:lim-bist}b).
At the turning points of the bistability curve a stable and an
unstable solution meet, so that the bifurcation is of the {\em saddle node}
type~\cite{hale}.

Our interest is however more focussed on instabilities of the second
kind, i.e. for $\kk_s\neq \kk_p$: this {\em parametric} instability signals
the onset of parametric oscillation with a finite intensity
appearing in a pair of distinct signal/idler modes at $\kk_{s,i}$.
From the point of view of bifurcation theory, the parametric instability
profoundly differs from the single-mode one.
As we shall see in the following, the pump-only solution still exists
beyond the threshold point, but it is no longer stable for an
eigenvalue of the linear stability matrix $L(\kk)$ has crossed the real
axis: the bifurcation is then of the {\em Hopf} type~\cite{hale} and is  
accompanied by a spontaneous breaking of a signal/idler $U(1)$ phase rotation 
symmetry~\cite{Goldstone}.

\subsection{Available range of signal/idler frequencies}
\label{sec:angle}

In the present paper, we shall not address the problem of the
determination of the wavevectors $\kk_{s,i}$ which are actually
selected by the parametric process above threshold.
This is a very complicate problem and is postponed to a
forthcoming publication~\cite{pattern}. 
Here we shall limit ourself to a study of the lower threshold for
parametric emission: the parametric oscillation dynamics will be
initiated as soon as the incident intensity exceeds the threshold
value for some pair of signal/idler modes.
 
For each value of pump wavevector $\kk_p$, it is important to
characterize the range of $\epsilon_{si}$ that can be obtained when
the signal/idler wavevectors $\kk_{s,i}$ are spanned through all
different polariton states: the search for the minimum value of the
threshold has in fact to be restricted to the region of $\epsilon_{si}$
values which are actually available.

\begin{figure}[tb]
\begin{center}
\includegraphics[width= \columnwidth,angle=0,clip]{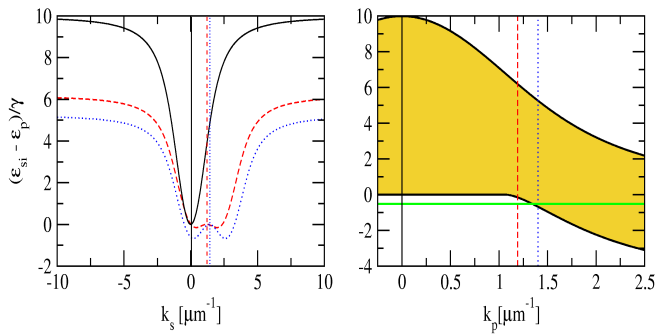}
\caption{
Left panel: plot of $\epsilon_{si}$ as a
function of $k_s$ for fixed values of $k_p=0$ (full black),
 $k_p=1.19\,\mu{\textrm  m}^{-1}$ (dashed red), 
$k_p=1.4\,\mu\textrm{m}^{-1}$ (dotted blue). 
The vertical lines show the value of the pump wave vector.
Right panel: band of available $\epsilon_{si}$ values as a function of
 $k_p$.
The horizontal green line shows the optimal detuning
$\Delta_{si,p}^{opt}$  (see Sec.\ref{sec:lowest} below), 
the vertical lines indicate the $k_p$ values
corresponding to the curves in the left panel.
 Cavity parameters as in Fig.\ref{fig:sketch}, equal damping rates
 $\gamma_{s,p,i}=\gamma$ with $\hbar \gamma=0.25$ meV.
}
\label{fig:pwsi}
\end{center}
\end{figure}

This point is addressed in Fig.\ref{fig:pwsi}. In the left panel, the 
behaviour of the detuning $(\epsilon_{si}-\epsilon_p)$ as a function 
of $k_s$ is shown for three different values of $k_p$ and the yellow region 
in the right panel summarizes the accessible detunings as a function of $k_p$.
For small $k_p$, the $\epsilon_{si}$ vs. $k_s$ curve has a single
minimum at $k_s=k_p$ where $\epsilon_{si}=\epsilon_p$, and then tends
to a finite limit for large $k_s$ ($\epsilon(k)$ has a finite limit
for large $k$).
For larger values of $k_p$, negative values
$\epsilon_{si}-\epsilon_p<0$ can be reached. 
The minimum is in fact split in two separate minima~\cite{whittaker05}, 
symmetrically located around the pump angle as required by the symmetry of
$\epsilon_{si}$ under exchange of the signal 
and idler modes.
The upper limit of the available band monotonically decreases as a
function of $k_p$, due to the corresponding increase of
$\epsilon_p$. In particular, it tends to $0$ for large values of
$k_p$.

\subsection{Pump intensity $n_{xp}$ at the parametric threshold}

As it often happens in nonlinear optical systems, it is useful to study
the parametric threshold first in terms of the internal light
intensity in the cavity, in our case the excitonic pump mode population
$n_{xp}$. Connection to the incident intensity $I_{inc}$ will
be then made in the next subsection.
As we are still left with several parameters, namely
$\gamma_i/\gamma_s$, $X_s^2$ and $X_i^2$, we are forced to restrict
the discussion to some illustrative examples. The qualitative features 
are however quite robust with respect to their variations.
Let us begin from the $\gamma_s=\gamma_i=\gamma_{si}$ case: as the
argument of the square root in eq.\eq{imlambda} is purely real, the
calculations are in this case the simplest.  

\begin{figure}[tb]
\begin{center}
\includegraphics[width= \columnwidth,angle=0,clip]{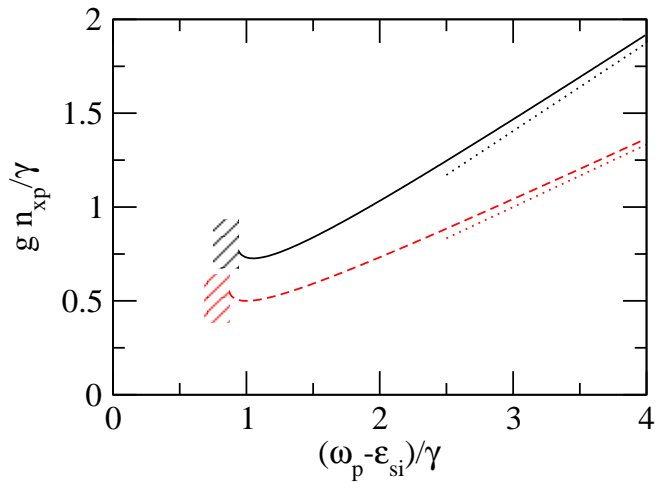}
\caption{
Threshold excitonic density in the pump mode as a function of the
pump laser detuning from the average signal/idler natural frequencies. 
The full black curve refers to a semiconductor microcavity for which the
Hopfield coefficients are $X_s=X(0)$ and $X_i=X(2k_{\rm magic})$.
The dashed red curve refers to the case of a cavity containing a passive
nonlinear material for which $X_j=1$.
The curves do not continue through the hatched region at the left hand
side where parametric oscillation can never take place.
The dotted lines represent the approximation \eq{p2threslarge}.
Cavity parameters as in Fig.\ref{fig:pwsi}. 
}
\label{fig:p2thres}
\end{center}
\end{figure}

The pump mode population $n_{xp}$ at the parametric
threshold is plotted
in Fig.\ref{fig:p2thres} as a function of $\omega_p-\epsilon_{si}$ for two
possible choices of the Hopfield coefficients. No qualitative
differences are visible, but only quantitative ones.
The main feature of these curves is the fact that parametric
oscillation can only take place for sufficiently large values of
$\omega_p-\epsilon_{si}$. The hatched regions indeed indicate
where parametric oscillation can never take place, no matter
how large the population of the pump mode is.
Remarkably, the minimum value of the threshold population is reached
just before the endpoint of the curves. 

Simple physical arguments can be put forward to explain these
features.
In a $\chi^{(3)}$ parametric oscillator, the nonlinearity not only
provides the parametric coupling between the signal and idler modes via the
off-diagonal terms in the matrix \eq{bog_matr}, but is at the same
time responsible for a mean-field blue shift of the signal and idler
mode frequencies by $2\,g\,|X_{s,i}|^2\,n_{xp}$. 
Once this shift is taken into account, the resonance condition for the
parametric process is renormalized to 
\begin{equation}
\omega_p=\epsilon_{si}+g(|X_s|^2+|X_i|^2)n_{xp}.
\label{resonance_P}
\end{equation}
From \eq{imlambda}, it is easy to see that the minimum value of
  the threshold
\begin{equation}
 n_{xp}^{min}=\frac{\gamma_{si}}{2g|X_s|\,|X_i|}
\label{P_min}
\end{equation}
is indeed attained when this condition is satisfied.
Combining \eq{resonance_P} and \eq{P_min} gives the optimal detuning
\begin{equation}
\omega_p-\epsilon_{si}=\gamma_{si}\,\frac{|X_s|^2+|X_i|^2}{2|X_s\,X_i|},
\label{opt_det}
\end{equation}
which corresponds to the position of the minimum of the curves plotted in
Fig.\ref{fig:p2thres}. 

On the other hand, for large and positive values of the detuning
$\omega_p-\epsilon_{si}\gg \gamma_{si}$, the threshold grows in a linear
way as a function of $\omega_p-\epsilon_{si}$
\begin{equation}
n_{xp}= \frac{\omega_p-\epsilon_{si}}{g(|X_s|^2+|X_i|^2+|X_s X_i|)}.
\label{p2threslarge}
\end{equation}

Finally, for $\omega_p-\epsilon_{si}<0$ the well-known inequality
$(|X_s|^2+|X_i|^2)^2 > |X_s|^2\,|X_i|^2$ implies that 
\eq{imlambda} can never be zero for any value of $n_{xp}$, 
so that parametric oscillation can never take place in this case.
The mean field shifts push in fact the signal/idler modes
out of resonance before the parametric coupling can overcome the
damping rate $\gamma_{si}$.

\subsection{Laser intensity at threshold}
\label{sec:laser_int}

\begin{figure}[tb]
\begin{center}\includegraphics[width=\columnwidth,angle=0,clip]{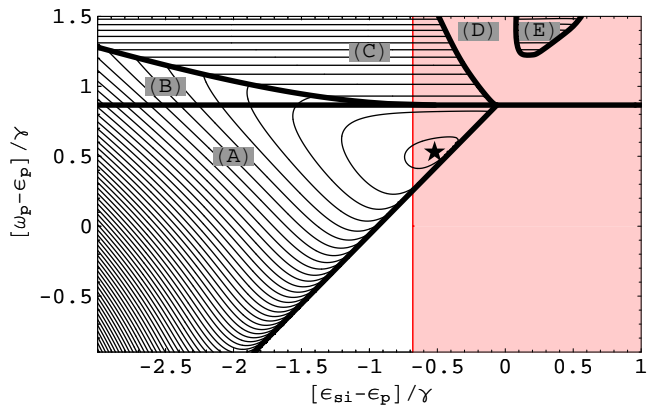}
\caption{
Contour plot of the threshold laser intensity as a function of the detunings.
The decay rates for the pump, signal and idler modes are taken
equal to $\gamma_{s,p,i}=\gamma$ and the
Hopfield coefficients $X_s=X(0)$, $X_p=X(k_{\rm
  magic})$ and $X_i=X(2k_{\rm magic})$. 
The lowest contour line in the plot is at $1.05 I_{inc}^{min}$,
where the minimum of the threshold intensity $I_{inc}^{min}$ is
 attained at the point indicated by a star $\star$
and is defined in
 Eq.\eq{I_inc_min} below.  
The difference between the contours is $0.3 I_{inc}^{min}$.
The letters (A)-(E) indicate the regions of qualitatively different
behaviours; the corresponding pump-only characteristic curves are
shown in Fig.\ref{fig:possibilities}.
The red-shaded area indicates the values of $\epsilon_{si}-\epsilon_p$
that are available for the value of the pump angle $k_p=1.4 \mu \rm m^{-1}$,
corresponding to the red curve in Fig.\ref{fig:pwsi}.
}
\label{fig:p2thr2D}
\end{center}
\end{figure}

\begin{figure}[htbp]
\begin{center}
\includegraphics[width=\columnwidth,angle=0,clip]{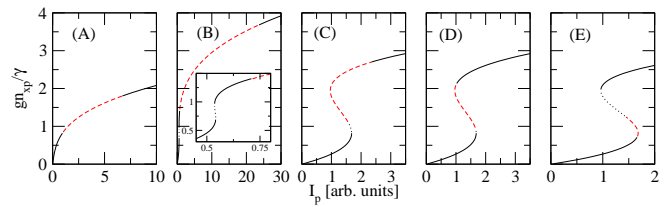}
\caption{Pump-only characteristic curves for different values of the
  detunings.
Instabilities with respect to a parametric oscillation process at a
  given $\epsilon_{si}$ are indicated as red dashed, pump-only
  instabilities are indicated as black dotted.
Arbitrary units for $I_p$ are the same in all panels.
The labels of the plots correspond to the different regions in
  Fig.\ref{fig:p2thr2D}: 
(A) parametric oscillation in the optical limiter regime 
($\epsilon_{si}-\epsilon_p=-1.5 \gamma,\;\omega_p-\epsilon_p=0$); ;
(B) bistability regime: parametric oscillation threshold above the
  pump-only bistability region 
($\epsilon_{si}-\epsilon_p=-2 \gamma,\;\omega_p-\epsilon_p=0.9 \gamma$), 
the inset shows the bistability region more in detail; 
(C) bistability regime: parametric oscillation threshold at pump-only
  bistability region 
($\epsilon_{si}-\epsilon_p=-0.6\gamma,\;\omega_p-\epsilon_p=1.5 \gamma$); 
(D) bistability regime: parametric oscillation threshold not reachable
  with upward ramp in laser  
intensity 
($\epsilon_{si}-\epsilon_p=-0.2 \gamma,\;\omega_p-\epsilon_p=1.5 \gamma$);
(E) bistability regime: parametric oscillation threshold before
  pump-only instability point 
($\epsilon_{si}-\epsilon_p=0.4 \gamma,\;\omega_p-\epsilon_p=1.5 \gamma$). 
}
\label{fig:possibilities}
\end{center}
\end{figure} 

In the previous section we have determined the value of the pump
mode population $n_{xp}$  
at the threshold for parametric oscillation.
The value of the corresponding laser intensity is then
obtained by using \eq{pump_only}. 
Care has to be paid to the fact that single-mode instabilities may make 
some branches of the bistability loop dynamically unstable and
therefore not reachable in an actual experiment.
Again, this feature is typical~\cite{whittaker05} of a $\chi^{(3)}$ OPO 
and is absent in $\chi^{(2)}$ ones, where the relation between the 
incident intensity and the pump mode population in the pump-only state 
is a linear one and no instability other than the parametric one is 
possible~\cite{lugiato}. 

Let us start from the $\gamma_p=\gamma_s=\gamma_i=\gamma$ case.
The predictions for the value of the laser intensity at the parametric 
threshold are summarized in Fig.\ref{fig:p2thr2D}, where the contour 
plot of the threshold laser intensity is shown as a function of the 
detuning $\epsilon_{si}-\epsilon_p$ between the signal/idler mode
frequencies and the pump mode frequency, and the detuning
$\omega_p-\epsilon_p$ of the pump laser from the pump mode frequency.
Throughout all the present discussion, the laser intensity is
assumed to be slowly but monotonically increased from $0$ until the
parametric threshold is reached.

The lower right corner of this figure corresponds to the hatched
region in Fig.\ref{fig:p2thres} where parametric oscillation can not
take place because $\omega_p$ is not sufficiently larger than
$\epsilon_{si}$. 

The heavy horizontal line at $\omega_p-\epsilon_p=\sqrt{3}\gamma/2$
separates the regions where the pump-only solution \eq{pump_only}
respectively shows optical limiting (below the line) and
optical bistability (above the line). 
In the {\em optical limiter} case shown in
Fig.\ref{fig:possibilities}A, the 
pump mode population $n_{xp}$ is a always a single valued function of
the pump laser intensity $I_p$. 
For a certain window in pump intensity, the (initially red-detuned)
signal/idler frequency $\epsilon_{si}$ is brought into resonance with
the pump energy $\omega_p$ by the mean-field shift, and the pump-only
state becomes unstable with respect to parametric oscillation 
(red dashed line).
Note that differently from the case of $\chi^{(2)}$ OPOs~\cite{lugiato},
parametric oscillation with $\chi^{(3)}$ media has an upper threshold
as well: for too large pump laser intensities, the blue-shift of the
signal/idler frequencies brings them off resonance and parametric
oscillation can no longer take place.

In the {\em optical bistability} case, the interplay between the
pump-only hysteresis with the parametric oscillation leads to a
variety of different behaviours (regions B-E).
In order to fully understand these issues, it is useful to identify
the relative position of the pump-only and the parametric instability
regions on the $n_{xp}$ vs. $I_p$ curves which are plotted in
Fig.\ref{fig:possibilities}. 
The different regions indicated in Fig.\ref{fig:p2thr2D} correspond in fact to
different arrangements of the two instability regions.

The simplest scenario is shown in Fig.\ref{fig:possibilities}B, where the
signal/idler frequency $\epsilon_{si}$ is very red-detuned from the
pump mode frequency $\epsilon_p$. 
The pump mode population needed to bring the signal/idler modes on
resonance is then much higher than the one needed to go through the
pump-only hysteresis loop. 
In this case, parametric oscillation occurs well above the bistability
region so that pump-only bistability and parametric oscillation are
effectively decoupled. The behaviour of parametric oscillation is
completely analogous to the optical limiter case. 

For the parameters of Fig.\ref{fig:possibilities}C, the pump only
instability still sets in before the parametric instability is
reached, but the state of the upper branch where the system is
expected to go, is parametrically unstable and OPO can start.
This means that the laser intensity threshold for parametric
oscillation coincides with the turning point of the hysteresis loop
and in particular no longer depends on the signal/idler frequency
$\epsilon_{si}$. For this reason, the contour lines shown in
region (C) of Fig.\ref{fig:p2thr2D} are straight horizontal lines. 

Fig.\ref{fig:possibilities}D shows a situation where the parametric
oscillation can not be reached by an upward ramp of the laser
intensity. 
For increasing pump laser intensity, the system jumps to the upper
branch of the hysteresis loop which is now parametrically stable in
the region of interest, so that parametric oscillation does not start.
Physically, the jumps shown by the pump mode population at the
switch-on point of the hysteresis loop is in fact large
enough to make the signal/idler detuning to jump directly from one side to the
other of the resonance.
Depending on the exact position of the parametric unstable region
along the hysteresis curve, parametric oscillation can possibly be
obtained by ramping the laser intensity down along the upper branch.
Finally, Fig.\ref{fig:possibilities}E describes the case when
parametric instability sets in before the bistability saddle node 
bifurcation is reached.

In Sec.\ref{sec:bif} we shall see that the parametric
instabilities shown in Fig.\ref{fig:possibilities}A-C lead to a
stable OPO state.
On the other hand, the situation is more complex for the case of
Fig.\ref{fig:possibilities}E, where it may happen that no stable
parametrically oscillating state is available and the system eventually
ends up in the upper branch of the pump-only hysteresis loop.

\subsection{Quest for the lowest threshold}
\label{sec:lowest}

In order to minimize the parametric threshold intensity, a
careful choice of the detunings has to be performed: in this section,
we will show that the mean-field shifts of the frequency modes make
this optimization problem somehow more complex than a trivial question
of ``magic angle''.

The optimal value of the detuning between pump frequency $\omega_p$ and
signal/idler frequency  
$\epsilon_{si}$ is given by \eq{opt_det}. 
In order to minimize the value of the incident pump intensity at
threshold, one has to simultaneously impose a resonance condition
between the pump laser frequency and the renormalized pump mode frequency:
\begin{equation}
\omega_p=\epsilon_p+g|X_p|^2n_{xp}.
\label{pump_renorm}
\end{equation}
The optimal pump and signal/idler mode detunings are then immediately
obtained by combining this result with \eq{P_min} and \eq{opt_det}: 
\begin{eqnarray}
\Delta_p^{opt}&=&\omega_p-\epsilon_p= \frac{\gamma}{2} \frac{|X_p|^2}{|X_s
 X_i|}.\\
\Delta_{si,p}^{opt}&=&\epsilon_{si} - \epsilon_p = -\frac{\gamma}{2} 
\frac{|X_s|^2+|X_i|^2-|X_p|^2}{|X_s X_i|}.
\label{opt_si}
\end{eqnarray}
The corresponding value of the threshold intensity is obtained by
simply substituting into Eq.~\eq{pump_only} and then using
Eq.~\eq{def:F_p}. 
For equal radiative and polaritonic decay rates $\gamma_{rad}=\gamma$,
one obtains:
\begin{equation}
I_{inc}^{min}= \frac{N_{tr} N_{QW} }{8 C_p^2 X_p^2 X_s X_i}
\frac{\hbar^2 \gamma^2 \omega_p}{\bar g},
\label{I_inc_min}
\end{equation}

In Fig.\ref{fig:p2thr2D}, the location of the minimum is indicated
by a star $\star$: in the present $\gamma_{s,p,i}=\gamma$ case, this point 
lies in the region (A) where the behaviour of the system is the simplest.
The pump-only solution being of the optical limiter type, no hysteresis
effects take place nor any interplay between parametric emission and
bistability .
Remarkably, both $\Delta_{si,p}^{opt}$ and $\Delta_{p}^{opt}$ and 
have a weak dependence on the Hopfield coefficients: for
$X_{s,p,i}=1$, they are equal to 
$\Delta_{si,p}^{opt}=-\gamma/2$, $\Delta_{p}^{opt}=\gamma/2$,
while they are approximately $\Delta_{si,p}^{opt}=-0.53 \gamma$,
$\Delta_{p}^{opt}=0.52 \gamma$, for the
typical values for a semiconductor microcavity used in
Fig.\ref{fig:p2thr2D}. 
This result is a refinement of the concept of ``magic angle'' at which
perfect resonance $\Delta_{si,p}=0$ is satisfied: as already noted in
Ref.\onlinecite{whittaker05}, 
small, but finite detunings $\Delta_{si,p}$ and $\Delta_{p}$ are
useful in a CW experiment to compensate the blue-shift of the 
signal/idler modes for increasing pump mode population.

The value \eq{opt_si} of the optimal detuning can be translated in the
wavevector space using the results of sec.\ref{sec:angle}: the
optimal detuning is indicated in Fig.\ref{fig:pwsi}b by the 
horizontal line. It is easy to see that this value can actually be
achieved as soon as $k_p>1.34 \mu \rm m^{-1}$, which corresponds for the
cavity parameters of Fig. \ref{fig:sketch} to a pump angle 
larger than $10$ degrees. 
This minimum pump angle depends on the damping rate $\gamma$: for a
smaller $\gamma$, the curves of Fig.\ref{fig:pwsi} are stretched in
the $y$-direction so that the optimal detuning can be already obtained
at smaller pump angles. 

The red-shaded area in Fig.\ref{fig:p2thr2D} indicates the values of
detuning $\Delta_{si,p}$ that are available for a pump wavevector
equal to $k_p=1.4\, \mu \rm m^{-1}$.
As the pump frequency $\omega_p$ can be chosen at will, no bound exist
in the vertical direction and  this area is bound only in the
horizontal direction.
For given values of $\omega_p$ and $k_p$, the parametric oscillation
dynamics is initiated when the incident pump intensity starts
exceeding the minimum value of the threshold on the horizontal segment
contained in the red-shaded area in Fig.\ref{fig:p2thr2D}.

A crucial role in the OPO operation is played by the damping rate.
From Eq.\eq{I_inc_min}, one sees that the laser intensity at the
optimal point is proportional to the square of the damping. 
Furthermore, the value of the damping affects the extent of the red
shaded area of available frequencies: for fixed pump angle, the border
of this area moves to the right upon increasing the damping rate and
eventually no longer overlaps with the (A) and (C) regions which are
most favourable for OPO operation (see Sec.\ref{sec:bif}).

\subsection{New features of the general
  $\gamma_s\neq\gamma_i\neq\gamma_p$ case} 
\label{sec:magic_angle}

\begin{figure}[htbp]
\begin{center}
\includegraphics[width= \columnwidth,angle=0,clip]{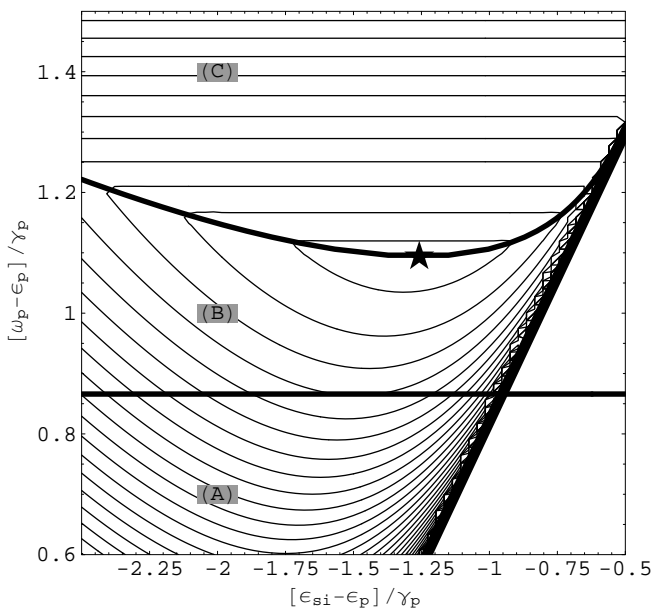}
\caption{
The same as Fig.\ref{fig:p2thr2D} for unequal damping rates
$\gamma_p= \gamma_s=\gamma_i/5$.
The lowest contour line in the plot is at $1.05 I_{inc}^{min}$,
where the minimum of the threshold intensity $I_{inc}^{min}$ is
 attained at the point indicated by a star $\star$.
The difference between the contours is $0.1 I_{inc}^{min}$.
}
\label{fig:p2thruneq2}
\end{center}
\end{figure} 

In current experimental configurations, e.g. for a pump in the vicinity 
of the ``magic angle'', the idler linewidth is often much larger than
the signal and pump ones, i.e. $\gamma_i\gg\gamma_p\simeq \gamma_s$.
Although the general formalism introduced in the previous sections
keeps holding its validity, some of the physical conclusions of
Sec.\ref{sec:lowest} have to be modified.
Because of the increased damping rate of the idler modes, the pump mode
population $n_{xp}$ required by \eq{imlambda} for the onset of 
the parametric oscillation corresponds to a blue-shift of the pump mode
$g |X_p|^2 n_{xp}$ much larger than $\gamma_p$.
With the frequency choice suggested by a naive application of the
condition \eq{pump_renorm}, the intensity value \eq{I_inc_min}
would correspond to the end-point of the upper branch of the bistability
curve. 
Unfortunately, this point can not be reached by the simple upward ramp 
of the pump laser intensity considered in the present paper, 
so that a more complete analysis is required which fully takes into 
account hysteresis effects. 

The results are shown in Fig.\ref{fig:p2thruneq2} for $\gamma_p=
\gamma_s=\gamma_i/5$. 
Because of the high value of the pump mode blue shift at the onset of
parametric oscillation, the (D) and (E) regions
are shifted to large values of $\Delta_p$ (not shown),
direct contact between the regions (A) and (C)
is lost and the gap is filled by the (B) region.
The optimal choice of the detunings lies on the border between the
regions (B) and (C): parametric oscillation starts on the upper branch
of the pump-only hysteresis curve exactly at the landing point of the
jump from the lower branch.
As one can see on Fig.\ref{fig:p2thruneq2}, the optimal
values of the $\Delta_{p,si}^{opt}$ and $\Delta_p^{opt}$ detunings
(measured in units of $\gamma_p$) are here larger than in the previous
$\gamma_{p,s,i}=\gamma$ case.  
On the other hand, the threshold intensity is increased above the naive
prediction \eq{I_inc_min} by a moderate factor of the order of two.

\subsection{Quantitative discussion \label{sec:quant_disc}}
\label{sec:quant}
In many technological applications of optical parametric oscillators, a
value as low as possible for the threshold intensity can be a key
advantage. 
In this respect, semiconductor microcavities in the strong coupling
regime are very promising systems thanks to the extremely high value
of the nonlinear coupling constant $g$, much higher than the one of
OPOs based on passive $\chi^{(3)}$ media (see discussion below Eq.
\eq{rel:g_chi3}). 

Using typical values  $\hbar \gamma=
0.1-0.5\,\textrm{meV}$ for the damping rates and $N_{QW}=3$ for the
number of quantum wells inside the cavity,
Eq.\eq{I_inc_min} yields a value in the $0.13-3.2 {\rm kW}/{\rm cm}^2$
range for the incident laser intensity at the parametric threshold, 
a value which is in rough agreement with experimental data of
Refs.~\onlinecite{stevenson,houdre,baumberg}.  
Given the scaling relation \eq{P_min}, the threshold intensity of
passive $\chi^{(3)}$ OPOs with a comparable quality factor is
orders of magnitude higher.  
Unless planar cavities of much higher quality factor are developed to
compensate for the much weaker nonlinearity of passive materials,
semiconductor microcavities in the strong coupling regime appear to be
most favourable systems in view of low-power OPO applications.  

In order for this comparison to be fair and complete, it is important
to extend it to the case of OPOs based on passive
$\chi^{(2)}$ materials~\cite{TROPO,lugiato,chi2OPO}. 
In this case, no mean-field shift of the mode frequencies occurs and
the minimum value of the parametric threshold is attained under the
resonance condition $\omega_p=\epsilon_{si}$ and is equal to:
\begin{equation}
|P|_{min}=\gamma/(2g_2),
\label{P_min_chi2}
\end{equation}
where the second-order nonlinear coupling constant for a planar cavity
of thickness $d$ and filled of a medium of linear dielectric constant
$\epsilon_{\rm lin}$ is
\begin{equation}
\hbar g_2=\mathcal C \chi^{(2)} \sqrt{\frac{(\hbar
    \omega_p)^3}{\epsilon^2_{\rm lin}d}}. 
\label{rel:g_chi2}
\end{equation}
At optimal pump detuning, the driving amplitude equals
$|F_p|=\gamma\,|P|/2$, irrespective of the type of nonlinearity. 
Combining this result with eqns. \eq{def:F_p}, \eq{rel:g_chi3},
\eq{P_min}, \eq{P_min_chi2} and \eq{rel:g_chi2}, one finds the
ratio between the threshold laser intensities of $\chi^{(2)}$ and
$\chi^{(3)}$ OPOs:
\begin{equation}
\frac{I^{(3)}_{min}}{I^{(2)}_{min}} =\frac{2\omega_p}{\gamma}
\frac{[\chi^{(2)}]^2}{\chi^{(3)}}.
\label{3/2}
\end{equation}
Using a typical value $\chi^{(3)}=10^{-9}\,\textrm{esu}$ for a
large Kerr nonlinearity, the value 
$\chi^{(2)}=4\times10^{-8}\,\textrm{esu}$ of the widely used KTP
crystal~\cite{NLO_mat}, and $\gamma/\omega_p=2\times 10^{-4}$ as in a
typical $\lambda/2$ semiconductor microcavity, the ratio \eq{3/2} turns
out to be around $0.016$.

This argument concludes the verification of the widespread expectation
that for comparable values of the quality factor, the threshold
intensity for parametric oscillation is orders of magnitude lower in
semiconductor microcavities in the strong coupling regime than in OPOs
based on passive $\chi^{(2,3)}$ materials.

\section{Bifurcation type and nonlinear solution above threshold
  \label{sec:bif}}

\begin{figure}[tb]
\begin{center}
\includegraphics[width= \columnwidth,angle=0,clip]{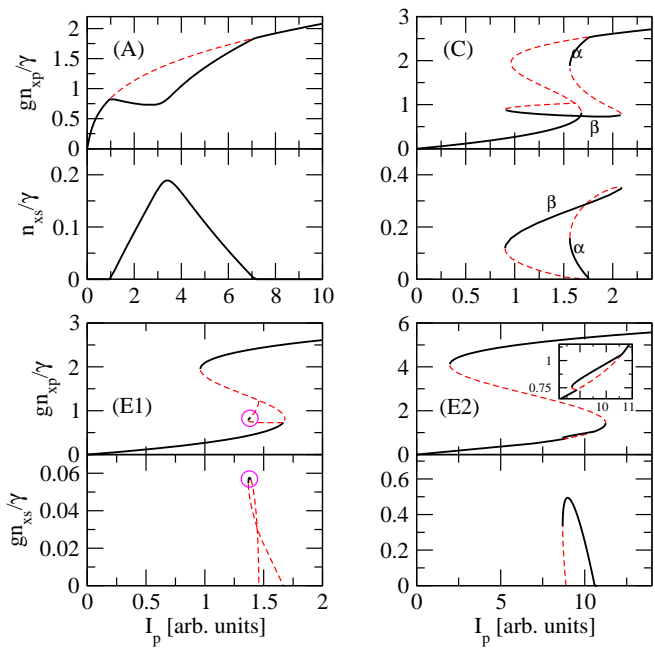}
\caption{
Pump and signal intensity as a function of the pump power in the
 different regimes. Black heavy full (red thin dashed) lines refer to
 the (un)stable solutions. 
$\gamma_{s,p,i}=\gamma$ are taken here, but the results are
 qualitatively robust to a change of the $\gamma$'s.
Arbitrary units for $I_p$ are the same in all
 panels.  
(A) Optical limiter regime
($\omega_p-\epsilon_p=0 , \epsilon_{si}-\epsilon_{p}=-1.5 \gamma$);
(C) Optical bistability regime where parametric and pump only threshold
 coincide 
($\omega_p-\epsilon_p=1.5 \gamma, \epsilon_{si}-\epsilon_{p}=-0.5 \gamma$);
(E1) Optical bistability regime where the parametric threshold precedes the 
pump only instability. The parametrically oscillating solution has a
very small stable part ($\omega_p-\epsilon_p=1.5 \gamma,
 \epsilon_{si}-\epsilon_{p}=0.4 \gamma$); 
(E2) Same as E1, with a different set of parameters such that
parametric oscillation is here possible for a wide range of pump intensities.
($\omega_p-\epsilon_p=3 \gamma, \epsilon_{si}-\epsilon_{p}=2 \gamma$). The inset
shows a magnification of the pump population in the OPO regime.
}
\label{fig:poss-bifurc}
\end{center}
\end{figure}

In the previous section, our attention has been focussed on the
behaviour of the system below the parametric threshold and we have
characterized the value of the threshold intensity as a function of
the detunings.
To complete the study, it is then necessary to investigate the
nature of the threshold point and characterize whether the onset of
parametric oscillation takes place in a continuous or discontinuous
way.
Both kind of behaviours have been indeed observed in the
experiments~\cite{dasbach,baumberg,baas}.
Correspondingly, the theoretical analysis in the present section will
show that a variety of different hysteresis effects can take place
depending on the kind of bifurcation that occurs at the threshold.
Again, our discussion here will be based on the assumption that
$\kk_{s,i}$ are  given quantities. A complete discussion of the
selection problem is postponed to a forthcoming
publication~\cite{pattern}. 

To make the analysis the simplest, a three-mode ansatz of the form:
\begin{equation}
\psi_{LP}(\kk,t) =S\,\delta_{\kk,\kk_s} e^{-i\omega _{s}t}
+P\, \delta_{\kk,\kk_p} e^{-i\omega_{p}t}
+I\, \delta_{\kk,\kk_i}e^{-i \omega_i t},
\label{psi0}
\end{equation}
can be used~\cite{whittaker01,whittaker05}, the signal/idler
frequencies and wave vectors being 
related by $\omega_i=2\omega_p-\omega_s$ and $\kk_i=2 \kk_p-\kk_s$.
By projecting the wave equation \eq{eq_mot} onto the three signal,
pump, and idler modes, the following equations of motion are found
\begin{widetext}
\begin{eqnarray}
i\frac{d}{dt}{\tilde P}&=&%
\left[ \epsilon_p-i\frac{\gamma}{2}-\omega _{p}\right] \tilde{P}+
g\,X_{p}^{2}\left[\left( \vert
\tilde{P}\vert ^{2}+2\vert \tilde{S}\vert ^{2}+2\vert \tilde{I}\vert
^{2}\right) \tilde{P}+2\tilde{P}^{\ast
}\tilde{S}\tilde{I}\right]+\tilde{F}_{p}
  \label{mf1} \\
i\frac{d}{dt}{\tilde S} &=&
\left[ \epsilon_s-i\frac{\gamma}{2}-\omega _{s}\right] \tilde{S}+
g\,X_{s}^{2}\left[\left( 2\vert
\tilde{P}\vert ^{2}+\vert \tilde{S}\vert ^{2}+2\vert \tilde{I}\vert
^{2}\right) \tilde{S}+\tilde{P}^{2}\tilde{I}^{\ast }\right]  \label{mf2} \\
i\frac{d}{dt}{\tilde I}&=&
\left[ \epsilon_{i}-i\frac{\gamma}{2}-2\omega _{p}+\omega
  _{s}\right] \tilde{I}+
g\,X_{i}^{2}\left[\left( 
2\vert \tilde{P}\vert ^{2}+2\vert \tilde{S}\vert ^{2}+\vert
\tilde{I}\vert ^{2}\right) \tilde{I}+\tilde{P}^{2}\tilde{S}^{\ast }\right],  \label{mf3}
\end{eqnarray}%
\end{widetext}
where the following shorthand notations have been introduced
$\epsilon_{p,s,i}=\epsilon(\kk_{p,s,i})$ and
$X_{p,s,i}=X(\kk_{p,s,i})$. 
Scaled quantities $\tilde{S}= X_{s}\, S$,  $\tilde{P}= X_{p}\, P$,
$\tilde{I}= X_{i}\, I$ and 
$\tilde{F}_{p}=X_p\,F_{p}$ have been also defined.
Imposing the stationarity of the solution and the condition that
$\omega_s$ is purely real gives a set of 7 real equations (3 complex
ones, plus 1 real equation) which has to be solved for a total of 8
real quantities: the three amplitudes ${\tilde S},{\tilde P},{\tilde
  I}$ and the (complex) 
parametric oscillation frequency $\omega_s$.
The extra degree of freedom which is left undetermined corresponds to
the 
 $U(1)$ signal/idler phase symmetry which is spontaneously broken
above the threshold~\cite{Goldstone}.

\begin{figure}[tb]
\begin{center}
\includegraphics[width= \columnwidth,angle=0,clip]{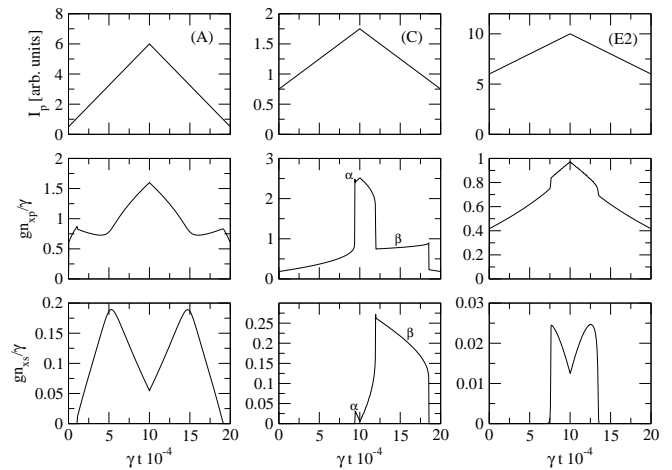}
\caption{
Time evolution of the pump laser intensity (upper panels), pump intensity (central panels)
and signal intensity (lower panels) for the same detuning parameters
as in the panels A, C and E2 of Fig.\ref{fig:poss-bifurc}.
}
\label{fig:dyn}
\end{center}
\end{figure}

Fig.\ref{fig:poss-bifurc} shows the behaviour of the pump $n_{xp}=|\tilde P|^2$ and
signal $n_{xs}=|\tilde S|^2$ mode occupations as a function of the incident pump
intensity $I_p$ for different choices of pump laser
$\omega_p-\epsilon_p$ and signal/idler detuning
$\epsilon_{si}-\epsilon_p$. 
These plots exemplify the system behaviour in the most significant
among the regimes studied in Fig.\ref{fig:possibilities}. Full lines
indicate stable regions, dashed lines are the unstable ones
\footnote{Stability has to be intended here within the three-mode
approximation: Eckhaus type instabilities due to the many modes in
which parametric oscillation can take place have not been taken into
account here and will be the subject of the forthcoming
publication~\cite{pattern}.}.
Correspondingly, a numerical integration of the time-dependent
equation of motion (\ref{mf1}-\ref{mf3})
 has been performed for a laser intensity which
is continuously swept up and down through the parametric threshold.
The resulting time-dependence of $n_{xp}$ and $n_{xs}$ is shown in
Fig.\ref{fig:dyn} for the most significant cases.

\subsection{Region (A): Optical limiter}

In the optical limiter case of Fig.\ref{fig:poss-bifurc}A, 
both the pump and the signal mode occupations are
continuous functions of the pump intensity.
The transition is analogous to a second-order phase transition: the
signal intensity is zero below and at the threshold and increases smoothly as a
function of the pump power.
In the language of nonlinear dynamics, this corresponds to a
so-called supercritical Hopf bifurcation~\cite{hale}. 
The corresponding time evolution is shown in the plots in the left
column of Fig.\ref{fig:dyn}: both $n_{xp}$ and $n_{xs}$ have a smooth
evolution in time which is immediately understood by following the
curves of Fig.\ref{fig:poss-bifurc}A.
The kinks correspond to the points where parametric emission switches
on and off.

It is interesting to compare this behaviour with the one of the
$\chi^{(2)}$ OPO in the $\Delta_0 \Delta_1 < 1$ regime discussed
in Ref.~\onlinecite{lugiato}. 
As one can see in Figs.\ref{fig:p2thr2D} and \ref{fig:p2thruneq2}, the
optical limiter regime corresponds to
$\epsilon_{si}-\epsilon_p<0$ and $\omega_p<\epsilon_p$. 
In both cases, the Hopf bifurcation is supercritical, and
the populations are continuous functions of the pump intensity across
the threshold. 
The behaviour  well above the threshold is however completely
different. In the $\chi^{(3)}$ case there is an upper threshold as
well, so that parametric oscillation disappears for too large a pump
intensity (not shown in the time-dependent plots). 
In the $\chi^{(2)}$ case, the parametric oscillation takes place
instead for all values of the laser intensity above the threshold. 
As shown in Ref.\onlinecite{lugiato},  for very high values of the incident
intensity it becomes however unstable towards self-pulsing and chaotic
behaviour.  

The behaviour of the system for the parameters of
Fig.\ref{fig:possibilities}B is completely analogous to the optical
limiter case: the pump only bistability and the parametric oscillation
indeed take place in an independent way.
In the OPO region, the behaviour of the pump and signal populations as
a function of the incident intensity is therefore closely analogous to
the one shown in Fig.\ref{fig:poss-bifurc}A.

\subsection{Region (C): Optical bistability}
The physics turns out to be much richer whenever parametric
oscillation and pump-only bistability take place in the same range of
intensities.
In the case shown in Fig.\ref{fig:possibilities}C, the pump-only
solution loses stability at the pump-only saddle node bifurcation. 
As the upper branch of the pump-only hysteresis loop is parametrically
unstable, the parametric oscillation sets in.
As shown in Fig.\ref{fig:poss-bifurc}C, the solution connecting the
lower and upper threshold for OPO can be a complicate
(multivalued) function~\cite{whittaker05}: typically, there are two
stable branches (indicated with $\alpha$ and $\beta$), which not
always can be reached in a continuous way by means of a simple upward
ramp of the pump intensity. 
To determine which branch is actually selected, the dynamics of the
system has to be considered (central column of Fig.\ref{fig:dyn}).
We have numerically found that the system jumps to the $\alpha$ branch as
soon as the pump intensity exceeds the pump-only turning point.
This $\alpha$ branch is then followed  during both the upward and the
following downward ramps until the saddle-node instability at the end 
of the branch is reached.
At this point the system has to jump to another solution: our
numerical simulations have shown that the $\beta$-branch is
dynamically selected, where parametric oscillation still takes place
with an even higher amplitude.
Finally, when the saddle-node instability at the end of
$\beta$ branch is reached, the system has no choice but to jump back
to the lower 
branch of the pump-only solution where parametric emission is no
longer present.

It is important to stress that this analysis is based on a three-mode
approximation: 
although this is certainly a valid description of a three-mode cavity,
it may be not representative of all that can happen in a many-mode
system such as a planar microcavity, where the $\alpha$ branch is
often Eckhaus-unstable against changes in the signal wavevector. 
On the other hand, the $\beta$ branch turns out to be generally much
more stable.
A more complete discussion of these issues will be presented in a
forthcoming publication~\cite{pattern}.

\subsection{Regions (E1,E2): Optical bistability \label{sec:E1E2}}
In the E region, the parametric instability occurs before the
pump-only one, and the corresponding Hopf bifurcation is generally of
the subcritical type~\cite{hale}. Two subcases are to be
distinguished. 

For $\epsilon_{si}\gtrsim \epsilon_p$ (Fig.\ref{fig:poss-bifurc}E1),
although the instability is of the parametric kind, no stable
parametrically oscillating solution exists for any pump intensity
above the threshold.
The parametric threshold is in fact very close to the pump-only
threshold, and only a very small part of the OPO solution is
stable (circle in Fig.\ref{fig:poss-bifurc}).
As this stable region corresponds to intensity values in between
the upper and lower turning points of the pump-only bistability loop,
parametric oscillation can not be reached by any continuous
intensity ramp, no matter its direction. 

For $\epsilon_{si} \gg \epsilon_p$ (Fig.\ref{fig:poss-bifurc}E2), the
parametric threshold is instead sufficiently lower than the pump-only
one for a stable OPO state to exist and to be reachable by means of an
upward intensity ramp:
a stable parametrically oscillating solution exists  in fact for laser
intensities extending from well below to well above the parametric
instability threshold.
However, as the bifurcation at the lower threshold is of the subcritical
Hopf type, parametric oscillation sets in in a discontinuous way for
an upward ramp in laser intensity.
A time-dependent calculation is then needed to ensure that the system
actually jumps from the lower branch of the pump-only hysteresis loop 
to the parametrically emitting solution rather than to
the upper branch of the pump-only bistability loop.
The results are shown in the right column of Fig.\ref{fig:dyn}:
the switch-on of the OPO emission during the upward ramp is
discontinuous, as well as the switch-off during the downward
ramp. 
This latter corresponds to a saddle-node instability at a pump
intensity slightly lower than the one of the subcritical Hopf
instability.
A new kind of hysteresis loop is therefore present: parametric
emission gives in fact a positive feedback to the pump mode population
and two solutions (a pump-only one and a parametrically emitting one)
are possible in a range of pump intensity values.
The main difference with respect to the standard pump-only hysteresis
loop is that the higher turning point is here at a Hopf bifurcation
rather than at a saddle-node one.

Remarkably, this phenomenology can be related to an analogous one shown by
a $\chi^{(2)}$ OPO in the $\Delta_0\,\Delta_1>1$ regime of
Ref.\onlinecite{lugiato}. 
Indeed,  $\omega_p>\epsilon_p$ and $\epsilon_{si} \gg \epsilon_p$ in
our (E2) region.
The qualitative shape of the parametrically-induced hysteresis loop
is indeed similar, with the main difference of the hysteresis loop
having a here a finite size also in the $n_{xp}$ vs. $I_p$ plot and not only in
the $P$ vs. $I_p$ one.
A qualitative analogy with the $\chi^{(2)}$ OPO can be found in
the (C) case as well: in addition to the topological similarity, 
the pump mode population is a very flat function of $I_p$ along the
$\beta$ branch, and the phase of the pump mode amplitude $P$ in the
$\beta$ branch differs from the one in the lower branch of the
bistability loop in a way very similar to the phase hysteresis shown in
the $\Delta_0\Delta_1<1$ case of Ref.\onlinecite{lugiato}.

\subsection{Considerations on quantum fluctuations}
  \label{sec:fluct} 

All the discussion so far has considered the polaritonic field as a
classical one, and therefore has neglected its fluctuations around the
mean-field value.
Before concluding, it is interesting to shortly address the behaviour of
the quantum fluctuations in the different cases.
The fluctuations around the pump-only solution below the threshold are
mostly determined by the nature of the instability at the threshold
point, i.e. whether this is a single-mode or a parametric one.
The physics of the fluctuations around the three-mode solution 
\eq{psi0} above the threshold is instead more complex~\cite{coh_length},
 and here we shall limit ourselves to a few, very general remarks. 

In regions (A) and (B), the onset of parametric oscillation
closely resembles a second-order phase transition: the signal, idler
and pump mode populations have a continuous dependence on the pump laser
intensity.
The overall behaviour as a function of the pump laser intensity is
qualitatively identical to the one discussed in Ref.\onlinecite{iac-prb} as a
function of the pump laser frequency: as the threshold point is
approached from below, the magnitude of the quantum fluctuations of the
signal and idler beam monotonically grows and eventually becomes very
large in the vicinity of the threshold where an eigenvalue of the
stability matrix \eq{bog_matr} goes to zero.
The fluctuations being due to parametric creation of signal-idler
polariton pairs, the signal and idler beams show significant quantum
correlation~\cite{ciuti_OPA,karr,langbein_QC}.
Above the threshold, the signal and idler fields have a finite mean-field
amplitude which continuously starts from zero.
Quantum fluctuations around this three-mode mean-field solution have a
more complex behaviour: a quite general fact is that the importance of
the fluctuations is most important close to the threshold point, and
then quickly decreases as one moves far from the threshold~\cite{coh_length}.

In the (E) cases, the behaviour is almost the same in the region below the
threshold:  
the instability having a parametric nature, the quantum fluctuations (as
well as the quantum correlations) in the signal and idler modes grow as
the threshold is approached and become strongest in the close vicinity
of the threshold point. 
On the other hand the behaviour above the threshold point is
dramatically different: the onset of the parametric oscillation
(provided it really starts, as in case E2) is discontinuous, and a
completely different solution branch is selected (Fig.\ref{fig:dyn}E2).
Furthermore, the landing point on the new branch is not necessarily in
the vicinity of the end-point of the branch, so fluctuations are
generally moderate. Yet, their magnitude becomes again large as one
approaches the end-point of the branch where one eigenvalue of the
stability matrix around the three-mode solution \eq{psi0} tends to zero.

In the (C) case, the behaviour is very different already below threshold:
as the instability at the end-point of the branch has a single-mode
nature, the quantum fluctuations in the signal and idler modes remain
moderate also in the vicinity of the threshold point, while the pump
mode ones grow very large as typical of optical bistable
systems~\cite{w-m}.

\section{Conclusions and outlook \label{sec:conclusions}}

In this paper we have given a systematic classification of the
behaviour of a triply resonant optical parametric oscillator based on
a semiconductor microcavity in the strong coupling regime.
Because of the $\chi^{(3)}$ nature of the collisional excitonic
nonlinearity, the interplay of optical bistability and optical
parametric 
oscillation makes the behaviour of these systems much richer
than the one of standard OPOs based on passive $\chi^{(2)}$ nonlinear
materials, and a variety of different threshold behaviours can be
found already within a simple three-mode theory. 
In agreement with recent experiments, depending on the specific value of the 
detunings, either a continuous switch-on or a
discontinuous jump can be found for the behavior of the signal intensity at the
parametric threshold.
The different behaviours have been classified by means of the general theory
of bifurcations, and a simple relation between the nature of the
instability point and the behaviour of the quantum fluctuations at the
threshold point has been pointed out.

In order to minimize the threshold incident intensity, a rigorous and
quantitative refinement of the ``magic angle'' criterion is provided
which takes into account the mean-field shift of the modes due to
interactions, as well as the possibility of hysteresis effects in the
pump-only dynamics.
A slight blue-detuning of the pump laser and a comparable red-detuning
of the signal/idler modes with respect to the pump mode frequency turns
out to be favourable in order to compensate for the mean-field shift of
the mode frequencies.

Generalization of the theory to the many-mode case is under way.
In order to take fully into account the inhomogeneous spatial
profile of the pump laser spot and the competition between parametric 
oscillation in different $k_s$ modes, techniques mutuated from the
theory of pattern formation  in nonlinear dynamical systems turn out to
be of great utility.

\acknowledgments

We are grateful to Cristiano Ciuti, Jer\^ome
Tignon, and Carole Diederichs for continuous stimulating discussions.
This research has been supported financially by the FWO-V projects Nos.
G.0435.03, G.0115.06 and the Special Research Fund of the University of
Antwerp, BOF NOI UA 2004.
M.W. acknowledges financial support from the FWO-Vlaanderen in the form
of a ``mandaat  Postdoctoraal Onderzoeker''.
We also acknowledge support by the Ministero dell'Istruzione,
dell'Universit\`a e della Ricerca (MIUR).

\end{document}